\pdfoutput=1
\documentclass[aps,prb,reprint,superscriptaddress,longbibliography]{revtex4-2}

\usepackage[T1]{fontenc}
\usepackage[utf8]{inputenc}
\usepackage{lmodern}
\usepackage{amsmath,amssymb}
\usepackage{graphicx}
\usepackage{bm}
\usepackage{placeins}
\usepackage[colorlinks=true,allcolors=blue]{hyperref}

\hypersetup{
  pdftitle={Three-Dimensional Fermiology and Thickness-Tuned Magnetotransport in Single-Crystalline Antimony Flakes},
  pdfauthor={Mikhail Gaponov, Jicheng Wang, Liang Zha, and Rui Wu}
}

\graphicspath{{figures/}}

\begin{document}

\title{Three-Dimensional Fermiology and Thickness-Tuned Magnetotransport in Single-Crystalline Antimony Flakes}

\author{Mikhail Gaponov}
\affiliation{Spin-X Institute, School of Physics and Optoelectronics, and State Key Laboratory of Luminescent Materials and Devices, South China University of Technology, Guangzhou 511442, China}
\author{Jicheng Wang}
\affiliation{Spin-X Institute, School of Physics and Optoelectronics, and State Key Laboratory of Luminescent Materials and Devices, South China University of Technology, Guangzhou 511442, China}
\author{Liang Zha}
\email{zhaliang@dhu.edu.cn}
\affiliation{College of Physics, Donghua University, Shanghai 201620, China}
\author{Rui Wu}
\email{ruiwu001@scut.edu.cn}
\affiliation{Spin-X Institute, School of Physics and Optoelectronics, and State Key Laboratory of Luminescent Materials and Devices, South China University of Technology, Guangzhou 511442, China}

\begin{abstract}
The extreme magnetoresistance of compensated semimetals is governed by both the Fermi-surface geometry and carrier relaxation, but these contributions are difficult to disentangle in finite-size structures. Here, we combine longitudinal and Hall magnetotransport measurements with temperature- and angle-dependent Shubnikov--de Haas oscillations in single-crystalline Sb flakes grown by chemical vapor deposition (CVD), with thicknesses ranging from 110 to 783~nm. As thickness increases, the non-saturating MR at 2~K and 14~T rises nearly 30-fold, reaching $7.13\times10^{5}\%$, while the primary frequency $F_{\alpha}$ remains approximately 99~T without any systematic shift. A joint three-channel analysis of $\rho_{xx}(B)$ and $\rho_{xy}(B)$ reveals that this evolution is driven by an increase in the mobility of a nearly compensated electron--hole pair, rather than by a reconstruction of the primary pockets. Angle-dependent measurements confirm the existence of a closed three-dimensional $\alpha$ pocket, and a reproducible high-frequency sector (335--377~T) is consistent with the electron $\beta$ orbit of the $L$-point pockets in bulk Sb. Together, the transport and quantum-oscillation results show that thickness tunes extreme MR through dimension-dependent scattering while preserving bulk-like fermiology.
\end{abstract}

\maketitle

\section{Introduction}

Magnetotransport in compensated semimetals is sensitive to both the electronic structure and momentum relaxation: low carrier concentrations combined with high electron and hole mobilities give rise to large, non-saturating magnetoresistance (MR) \cite{ref1}. Exceptionally large orbital responses in high-mobility systems are commonly referred to as extreme magnetoresistance (XMR). XMR has been reported in WTe$_2$ and NbP \cite{ref2,ref3}, rare-earth monopnictides including LaSb \cite{ref4,ref24}, as well as Cd$_3$As$_2$ and the diphosphides WP$_2$/MoP$_2$ \cite{ref25,ref26}. Its observation in elemental gray arsenic \cite{ref30} extends this behavior to elemental semimetals. Although XMR is often discussed alongside topologically nontrivial band structures, comparative studies show that it is not a unique signature of topology: near electron--hole compensation, high carrier mobility, and anisotropic multiband transport can generate an equally strong response within a semiclassical framework \cite{ref4,ref8,ref22,ref27,ref28,ref29,ref31}. These findings also show that $\rho_{xx}(B)$ alone cannot distinguish between topological, compensation-driven, and multiband mechanisms. Consequently, a joint analysis of longitudinal and Hall responses, supplemented by independent information on the Fermi surface, is required.

Antimony (Sb) serves as an ideal model for such an investigation. Its rhombohedral A7 structure (trigonal space group $R\overline{3}m$, No.~166), small band overlap, strong spin--orbit coupling, and multivalley anisotropic Fermi surface provide large non-saturating MR without the need to invoke intrinsic topological mechanisms \cite{ref5,ref6,ref7,ref8}.

In thin samples, the interplay between the mean free path and the geometric dimension becomes critical. In pure Sb, the transport mean free path $\ell_{\mathrm{tr}}$ can reach mesoscopic scales \cite{ref9,ref10}. The presence of metallic surface states on Sb(111) introduces additional complexity \cite{ref11,ref12}, yet their existence does not inherently imply the dominance of surface conduction. When $\ell_{\mathrm{tr}}$ becomes comparable to the thickness $t$, boundary scattering modifies the transport mobility even in the absence of significant reconstruction of the quantum orbits \cite{ref13,ref14}.

Thickness dependence, therefore, can manifest in two distinct regimes. In atomically thin Sb, quantum confinement and structural modifications can alter the electronic spectrum \cite{ref15,ref16}, whereas in thicker films and flakes, the classical size effect becomes dominant \cite{ref13,ref14,ref17}. The former should lead to shifts in extremal areas and Shubnikov--de Haas (SdH) frequencies, while the latter involves changes in carrier mobility $\mu$ and MR with constant frequencies. Angle-dependent measurements further distinguish between two-dimensional and closed three-dimensional orbits.

The mesoscopic thickness range of several hundred nanometers is particularly informative: the band structure remains bulk-like, while transport can already be surface-limited. In this regime, thickness acts as a controlled parameter to decouple changes in the energy spectrum from changes in scattering.

We use the classical magnetotransport tensor to determine compensation and transport relaxation times, and SdH oscillations to probe extremal orbits and quantum broadening \cite{ref18,ref19,ref20}. Together, these measurements distinguish spectral modifications from scattering effects.

CVD-grown single-crystalline Sb flakes with thicknesses from 110 to 783~nm exhibit dimension-dependent scattering while maintaining bulk-like fermiology. The primary SdH frequency $F_{\alpha}$ remains approximately 99~T and exhibits a three-dimensional angular dependence, while the high-frequency sector is consistent with the electron $\beta$ orbit of the $L$-point pockets in bulk Sb. We find $\mu_{\mathrm{pair}}\propto t^{0.87}$ and $\mathrm{MR}(14~\mathrm{T})\propto\mu_{\mathrm{pair}}^{1.99}$. An effective transport scale constructed from $\mu_{\mathrm{pair}}$ and $k_{\alpha}$ remains on the order of several sample thicknesses, consistent with dimension-dependent transport relaxation.

\section{Experiment}

\subsection{Sample Preparation and Structural Characterization}

Single-crystalline Sb flakes were grown via CVD on mica, transferred onto SiO$_2$/Si substrates using a polymer-assisted method, and patterned into multi-terminal devices with Ti/Au contacts.

\begin{figure*}[t]
  \centering
  \includegraphics[width=0.90\textwidth]{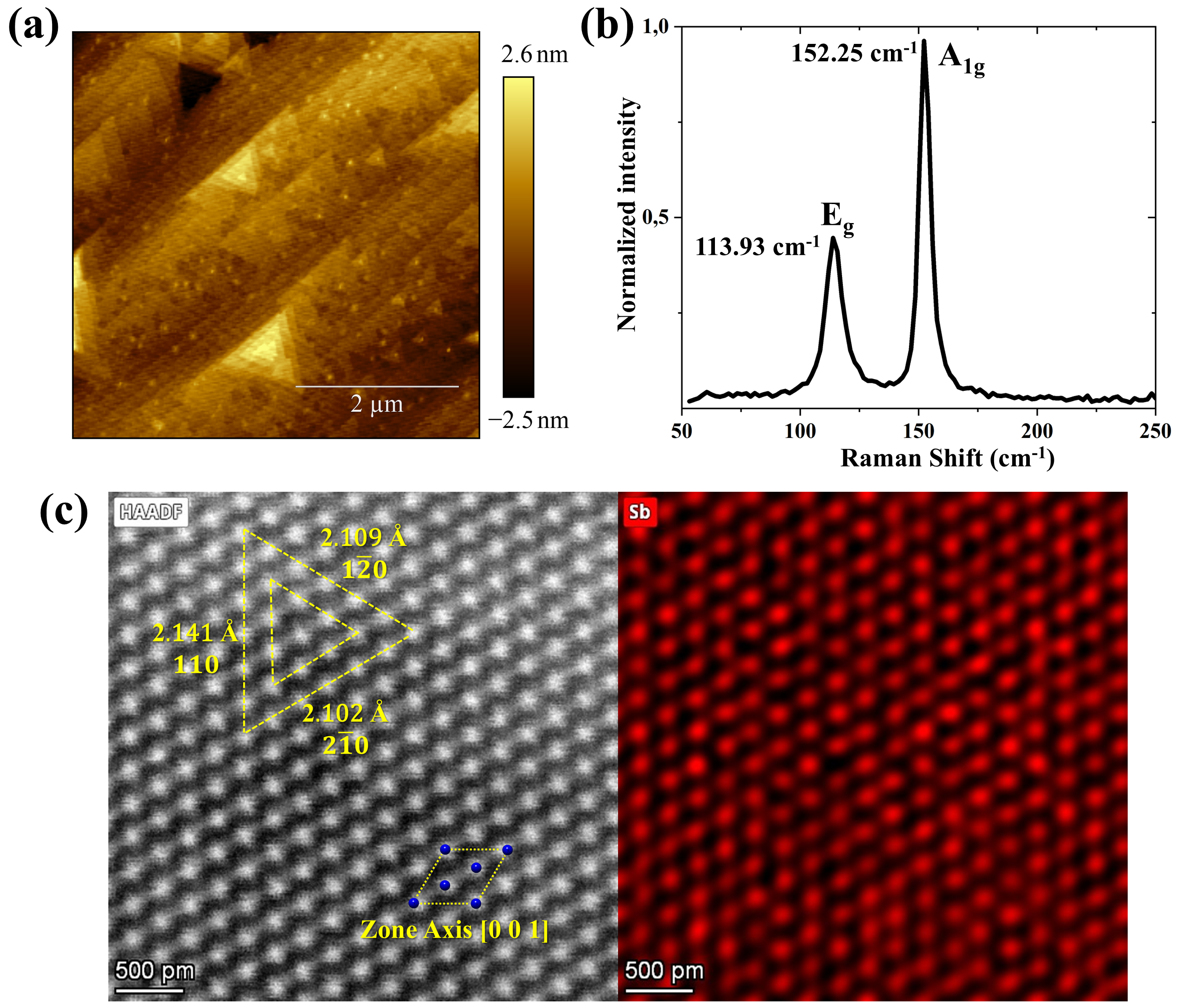}
  \caption{Structural characterization of CVD-grown Sb flakes. (a) Representative atomic force microscopy (AFM) topography of a triangular Sb flake; the color scale denotes surface height. (b) Raman spectrum measured from an Sb flake showing the $E_g$ and $A_{1g}$ modes. (c) Left: atomic-resolution HAADF-STEM image acquired along the [001] zone axis, revealing the ordered projection of the Sb lattice. The measured lattice spacings along three symmetry-equivalent in-plane directions are 2.141, 2.109, and 2.102~\AA, indexed to the (110), $(1\overline{2}0)$, and $(2\overline{1}0)$ planes, respectively. The projected lattice motif is highlighted by blue spheres and yellow dashed lines. Right: corresponding atomic-resolution Sb elemental map, showing that the Sb signal reproduces the periodic atomic-column arrangement observed in the HAADF-STEM image.}
  \label{fig:structure}
\end{figure*}

Figure~\ref{fig:structure} summarizes the morphology and structural characterization of the CVD-grown material. Atomic-force microscopy (AFM) reveals triangular Sb domains [Fig.~\ref{fig:structure}(a)]. The Raman spectrum exhibits the characteristic $E_g$ and $A_{1g}$ modes at 113.93 and 152.25~cm$^{-1}$, respectively, consistent with the A7 phase of Sb \cite{ref16} [Fig.~\ref{fig:structure}(b)]. Atomic-resolution high-angle annular dark-field scanning transmission electron microscopy (HAADF-STEM) viewed along [001] resolves the ordered Sb lattice and three symmetry-equivalent in-plane spacings of 2.102--2.141~\AA; the corresponding Sb elemental map reproduces the atomic-column arrangement [Fig.~\ref{fig:structure}(c)].

\begin{figure*}[t]
  \centering
  \includegraphics[width=0.95\textwidth]{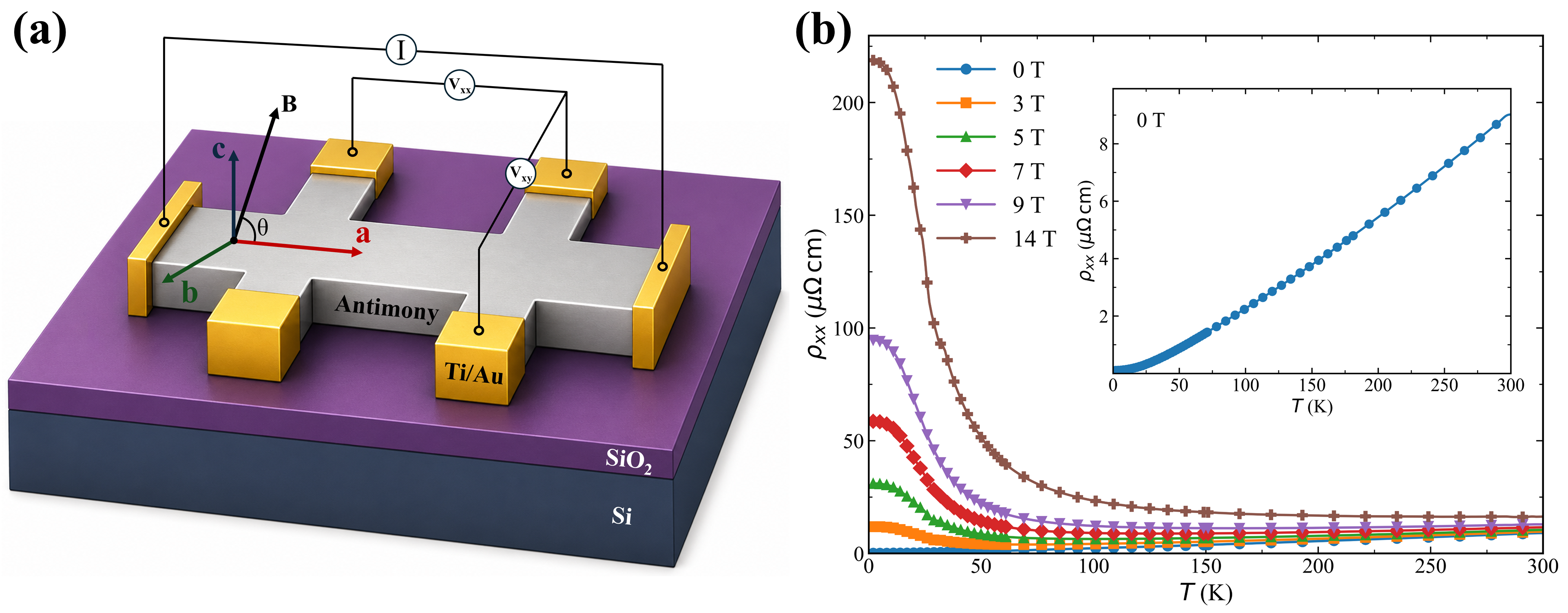}
  \caption{Device geometry and temperature-dependent magnetotransport. (a) Schematic of the multiterminal Sb device on Si/SiO$_2$ with Ti/Au contacts, defining the current and voltage configurations, crystallographic axes, magnetic field $B$, and angle $\theta$ measured from the flake plane. (b) Temperature dependence of the longitudinal resistivity of the Sb flake in a transverse magnetic field; the inset shows the zero-field curve.}
  \label{fig:rt}
\end{figure*}

Longitudinal and Hall voltages were measured using the four-probe configuration shown in Fig.~\ref{fig:rt}(a), at a current of 100~$\mu$A, corresponding to nominal average current densities of $1.70\times10^{3}$--$7.58\times10^{3}$~A\,cm$^{-2}$ across the four devices. Measurements were performed in an Oxford Instruments TeslatronPT system in magnetic fields up to 14~T and over a temperature range of 2--300~K. For the analysis of SdH oscillations, data acquired at temperatures up to 30~K were used.

\subsection{Transport Measurement Geometry}

The flakes expose the basal (001) plane of the A7 lattice, with the trigonal $c\parallel[001]$ axis normal to the surface [Fig.~\ref{fig:rt}(a)]. The crystallographic $a$ and $b$ axes lie in the basal plane and subtend $120^{\circ}$. For transport, we use an orthogonal experimental frame: $C_1\parallel I\parallel a$, $C_2$ is the transverse in-plane direction, and $C_3\parallel c\parallel[001]$. Thus, $C_2$ is not identified with the non-orthogonal crystallographic $b$ axis. The angle $\theta$ is measured from the flake plane, as shown in Fig.~\ref{fig:rt}(a): $\theta=90^{\circ}$ corresponds to $B\parallel C_3$ and $\theta=0^{\circ}$ to an in-plane field. Measurements were performed in the $C_1$--$C_3$ and $C_2$--$C_3$ rotation planes, for which the in-plane field projection was parallel and perpendicular to the current, respectively.

\section{Results}

\subsection{Temperature Dependence of Resistivity}

Longitudinal resistivity was measured from 2 to 300~K in magnetic fields of 0, 3, 5, 7, 9, and 14~T with $B\parallel C_3$ [Fig.~\ref{fig:rt}(b)]. Resistivity values were extracted from the measured resistance using the channel geometry:
\begin{equation}
\rho_{xx}=R_{xx}\frac{wt}{L},
\label{eq:resistivity}
\end{equation}
where $L$ is the distance between the longitudinal probes, $w$ is the channel width, and $t$ is the flake thickness. All $\rho_{xx}$ values are presented in $\mu\Omega\,$cm.

In zero field, the flake exhibits metallic behavior: $\rho_{xx}$ decreases monotonically from 9.02~$\mu\Omega\,$cm at 300~K to 0.109~$\mu\Omega\,$cm at 2~K, with $d\rho_{xx}/dT$ remaining positive throughout the temperature range [inset of Fig.~\ref{fig:rt}(b)]. The corresponding residual resistance ratio (RRR) is 82.7, calculated as
\begin{equation}
\mathrm{RRR}=\frac{\rho_{xx}(300~\mathrm{K})}{\rho_{xx}(2~\mathrm{K})}.
\label{eq:rrr}
\end{equation}
Across the four flakes, the RRR increases systematically with thickness from 15.68 to 114.4. For comparison, an RRR of approximately 350 was reported for macroscopic Sb \cite{ref8}, while values ranging from approximately 250 to 4200 were obtained for high-purity bulk Sb crystals \cite{ref9}. The lower values in the flakes and their systematic increase with thickness are consistent with the stronger contribution of surface and boundary scattering in thinner samples. Nevertheless, the large RRR of the thickest flake demonstrates that low residual scattering and high electronic quality are retained after CVD growth, transfer, and lithographic processing.

Upon application of a magnetic field, a characteristic resistivity upturn (turn-on), typical of compensated high-mobility semimetals, is observed. The minimum in $\rho_{xx}(T)$ shifts from $T^{*}\approx71$~K at 3~T to approximately 288~K at 14~T, while $\rho_{xx}(2~\mathrm{K})$ increases from 11.86 to 218.8~$\mu\Omega\,$cm. The zero-field resistivity remains metallic.

This temperature dependence of $\rho_{xx}(T,B)$ reflects two competing trends: cooling lowers the zero-field resistivity $\rho_0(T)$ but increases carrier relaxation times and mobilities, enhancing the orbital response. In the limit of a nearly compensated two-band semimetal, this relationship is qualitatively described by \cite{ref23,ref24,ref27,ref28}
\begin{equation}
\rho_{xx}(T,B)\approx\rho_0(T)\left[1+\mu_e(T)\mu_h(T)B^2\right].
\label{eq:two_band_rt}
\end{equation}
At high temperatures, the mobilities are small, and the magnetic field has a negligible effect on the resistivity. Below $T^{*}$, the increase in the product $\mu_e\mu_h$ outweighs the decrease in $\rho_0$, causing a sign change in $d\rho_{xx}/dT$; at even lower temperatures, $\rho_{xx}$ reaches a plateau determined by residual scattering. This phenomenon is an orbital manifestation of carrier compensation, rather than a transition into an insulating state \cite{ref2,ref22,ref23,ref24,ref27,ref28}.

\subsection{Magnetoresistance and Its Thickness Dependence}

Comparing flakes of different thicknesses allows us to examine the role of boundary scattering while keeping the chemical composition, growth method, and field orientation fixed.

\begin{figure*}[t]
	\centering
	\includegraphics[width=\textwidth]{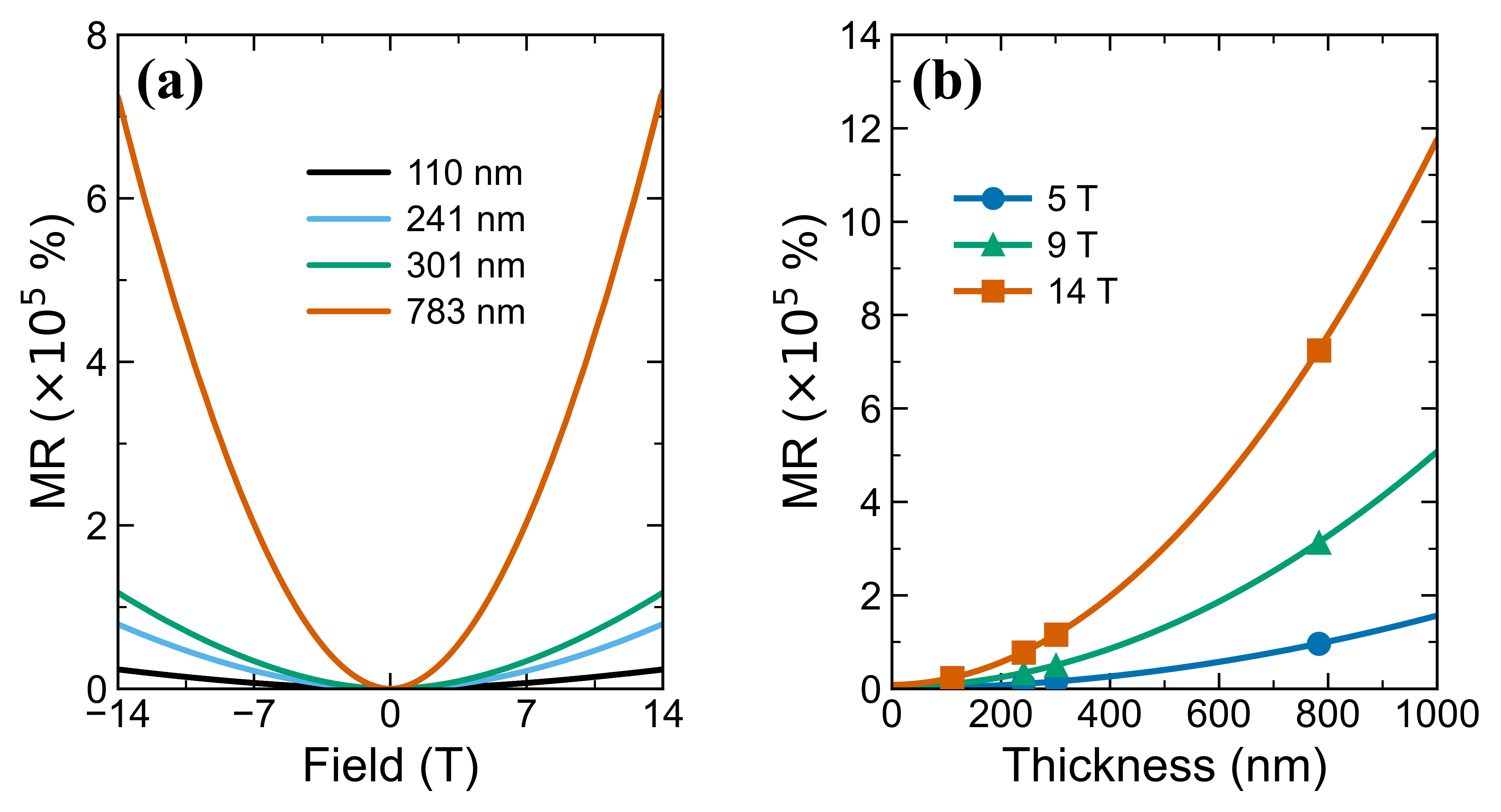}
	\caption{(a) Longitudinal magnetoresistance of Sb flakes with different thicknesses. (b) Thickness dependence of magnetoresistance. Symbols represent experimental data, and solid lines show fits to $\mathrm{MR}\propto B^2$.}
	\label{fig:mr}
\end{figure*}

MR of Sb flakes with different thicknesses was measured at 2~K, a regime where phonon scattering is minimized, allowing the differences in residual mobility to be most clearly observed. For all samples, the magnetic field was oriented perpendicular to the surface ($B\parallel C_3$). The relative magnetoresistance was defined as
\begin{equation}
\mathrm{MR}(B)=\frac{\rho_{xx}(B)-\rho_{xx}(0)}{\rho_{xx}(0)}\times100\%.
\label{eq:mr}
\end{equation}
Here, $\rho_{xx}(0)$ is the zero-field resistivity for each individual device. The field-dependent curves, shown in Fig.~\ref{fig:mr}(a), are symmetric in $B$ and exhibit non-saturating behavior up to 14~T. Reducing the thickness down to 110~nm significantly suppresses the MR amplitude but does not alter its fundamental orbital character.

In the investigated field range (2--14~T), the field dependence follows a power law:
\begin{equation}
\mathrm{MR}(B)=C|B|^m,
\label{eq:mr_power}
\end{equation}
where $m\approx1.8$--1.9 (specifically, $m=1.76$, 1.90, 1.84, and 1.88 for 110, 241, 301, and 783~nm, respectively; $R^2>0.9995$). The exponent $m$ is close to 2 and exhibits no monotonic trend with thickness, implying that the thickness primarily modulates the scale of the orbital MR rather than its underlying physical mechanism.

At 14~T, the MR values are $2.37\times10^4\%$, $7.89\times10^4\%$, $1.17\times10^5\%$, and $7.13\times10^5\%$ for the 110, 241, 301, and 783~nm samples, respectively. Even the 110-nm flake maintains a giant non-saturating response, while the 783-nm device reaches the extreme-magnetoresistance levels typical of bulk semimetals. Increasing the thickness from 110 to 783~nm (a 7.1-fold increase) enhances $\mathrm{MR}(14~\mathrm{T})$ by approximately 30-fold, a trend that is consistent at 5 and 9~T [Fig.~\ref{fig:mr}(b)]. The data for all three fields obey a common scaling law:
\begin{equation}
\mathrm{MR}(B,t)=A(B)t^{p(B)}.
\label{eq:mr_thickness}
\end{equation}
Logarithmic fitting provides exponents $p=1.70$, 1.73, and 1.74 at 5, 9, and 14~T ($R^2=0.993$--0.996). The ratio of the MR amplitudes between thick and thin flakes is nearly field-independent, remaining between 27.6 and 30.0.

In the nearly compensated limit, $\mathrm{MR}\approx\mu_e\mu_hB^2$ \cite{ref26,ref27,ref28,ref31}. Consequently, the observed quasi-quadratic field dependence, accompanied by a large change in amplitude, indicates that mobility is the primary thickness-dependent parameter. Decreasing $t$ enhances surface scattering, whereas increasing $t$ reduces the relative role of boundary scattering, as expected in the Fuchs--Sondheimer picture \cite{ref13,ref14}.

\subsection{Joint Multiband Analysis, Compensation, and Carrier Mobilities}

The observed quasi-quadratic MR behavior indicates a strong orbital response but does not inherently disentangle the contributions of carrier concentration and mobility. In multiband semimetals, the longitudinal conductivity is the sum of all channel contributions, whereas the Hall component is determined by their signed differences and is more heavily weighted toward high-mobility carriers. We therefore fit $\rho_{xx}(B)$ and $\rho_{xy}(B)$ simultaneously using a single set of parameters \cite{ref1,ref4,ref8,ref22,ref23,ref27,ref28,ref29}.

Due to unavoidable geometric asymmetries in the contact placement, longitudinal and Hall voltages exhibit mutual mixing. To eliminate this effect, the experimental data for both voltage components were decomposed into even and odd symmetries with respect to the magnetic field:
\begin{align}
\rho_{xx,\mathrm{even}}(B)&=\frac{\rho_{xx}(B)+\rho_{xx}(-B)}{2},
\label{eq:even}\\
\rho_{xy,\mathrm{odd}}(B)&=\frac{\rho_{xy}(B)-\rho_{xy}(-B)}{2}.
\label{eq:odd}
\end{align}
This even--odd decomposition removes the Hall admixture from the longitudinal signal and the magnetoresistive admixture from the Hall signal. All datasets were analyzed within a single field range of $|B|\leq14$~T.

We describe the experimental curves using a semiclassical three-channel model consisting of two electron-like channels ($e_1$, $e_2$) and one hole-like channel ($h$). For each channel, defined by carrier concentration $n_i$, scalar mobility $\mu_i$, and sign $s_i$ ($-1$ for electrons and $+1$ for holes), the conductivity components were expressed as \cite{ref1,ref4,ref28}
\begin{align}
\sigma_{xx}(B)&=\sum_i\frac{e n_i\mu_i}{1+(\mu_iB)^2},
\label{eq:sigma_xx}\\
\sigma_{xy}(B)&=\sum_i\frac{e s_i n_i\mu_i^2B}{1+(\mu_iB)^2}.
\label{eq:sigma_xy}
\end{align}
The fitting was performed after inverting the full conductivity tensor:
\begin{equation}
\rho_{xx}=\frac{\sigma_{xx}}{\sigma_{xx}^2+\sigma_{xy}^2},\qquad
\rho_{xy}=\frac{-\sigma_{xy}}{\sigma_{xx}^2+\sigma_{xy}^2}.
\label{eq:rho_tensor}
\end{equation}

A single set of six parameters was used to fit both components simultaneously. To ensure stability, the optimization was performed in the logarithmic space of $n_i$ and $\mu_i$ with positive constraints. During the final optimization, the weights of $\rho_{xx}(B)$ and $\rho_{xy}(B)$ were balanced to account for their significantly different dynamic ranges, and the values at $B=0$ and $|B|=14$~T were treated as constrained anchors to prevent the high-field data from dominating the fit. We found that using independent parameter sets for each thickness provided a superior description of the experimental evolution compared to models with constrained carrier concentrations.

\begin{figure*}[t]
  \centering
  \includegraphics[width=\textwidth]{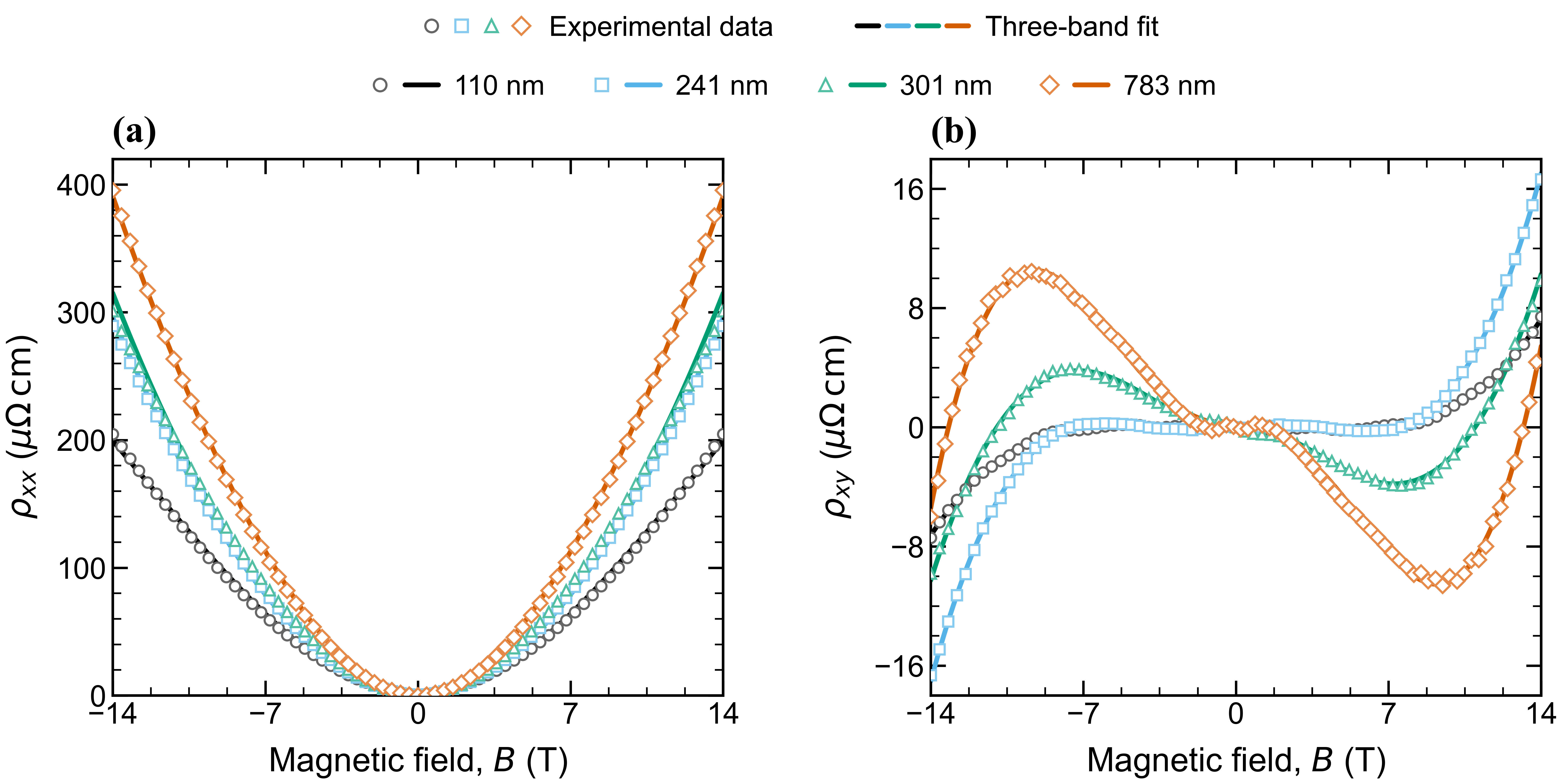}
  \caption{Joint three-channel fitting of (a) $\rho_{xx,\mathrm{even}}$ and (b) $\rho_{xy,\mathrm{odd}}$ for Sb flakes with different thicknesses.}
  \label{fig:threeband}
\end{figure*}

With a single set of parameters for each flake, the model reproduces the longitudinal MR and nonlinear Hall response across the thickness series, including the measured zero-field $\rho_{xx}(0)$ and high-field value $\rho_{xx}(14~\mathrm{T})$. Because the objective function uses nonuniform weights for the two components and field ranges, we assess the fits using the residuals of both components and agreement with the measured low- and high-field values, rather than a single normalized root-mean-square error [Fig.~\ref{fig:threeband}].

The resulting parameters (Table~\ref{tab:threeband}) exhibit a consistent structure across all flakes: the $e_1$ and $h$ channels constitute a high-mobility electron--hole pair, while the $e_2$ channel is characterized by significantly lower mobility. To quantify the compensation and the characteristic mobility scale of this high-mobility pair, we define
\begin{align}
\eta_{\mathrm{pair}}&=\frac{2|n_h-n_{e1}|}{n_h+n_{e1}},
\label{eq:eta_pair}\\
\mu_{\mathrm{pair}}&=\sqrt{\mu_{e1}\mu_h}.
\label{eq:mu_pair}
\end{align}
The $\eta_{\mathrm{pair}}$ values are 0.041\%, 0.339\%, 1.643\%, and 0.555\% for thicknesses of 110, 241, 301, and 783~nm, respectively. This confirms that nearly perfect compensation of the $e_1$--$h$ pair is maintained across the entire thickness range. The characteristic mobility $\mu_{\mathrm{pair}}$ scales as $t^{0.87}$ ($R_{\log}^2\approx0.990$), which is consistent with the observed growth in MR.

\begin{table*}[t]
\caption{Effective carrier concentrations $n_i$ and mobilities $\mu_i$ from the joint fit of $\rho_{xx}$ and $\rho_{xy}$ at 2~K. The $e_1$ and $e_2$ channels are electron-like; $h$ is hole-like.}
\label{tab:threeband}
\small
\setlength{\tabcolsep}{1.7pt}
\begin{ruledtabular}
\begin{tabular}{c c c c c c c}
$t$ (nm) & $n_{e1}$ ($10^{20}$~cm$^{-3}$) & $\mu_{e1}$ ($10^3$~cm$^2$\,V$^{-1}$\,s$^{-1}$) & $n_{e2}$ ($10^{20}$~cm$^{-3}$) & $\mu_{e2}$ ($10^3$~cm$^2$\,V$^{-1}$\,s$^{-1}$) & $n_h$ ($10^{20}$~cm$^{-3}$) & $\mu_h$ ($10^3$~cm$^2$\,V$^{-1}$\,s$^{-1}$) \\
\colrule
110 & 2.926 & 12.192 & 0.492 & 0.135 & 2.928 & 12.607 \\
241 & 3.953 & 22.168 & 0.078 & 0.577 & 3.967 & 20.884 \\
301 & 4.533 & 23.151 & 0.106 & 1.506 & 4.608 & 30.084 \\
783 & 7.904 & 50.893 & 0.095 & 0.721 & 7.948 & 90.915 \\
\end{tabular}
\end{ruledtabular}
\end{table*}

The physical role of the $e_2$ channel is manifested only when considering the full magnetic-field range. Its contribution to the longitudinal conductivity,
\begin{equation}
w_{e2}(B)=\frac{\sigma_{xx,e2}(B)}{\sum_i\sigma_{xx,i}(B)},
\label{eq:e2_weight}
\end{equation}
is only 0.092\%, 0.0265\%, 0.0657\%, and 0.0061\% at $B=0$ for thicknesses of 110, 241, 301, and 783~nm, respectively. However, as the high-mobility pair is rapidly suppressed in high fields, the $e_2$ contribution increases to 21.1\%, 12.7\%, 14.2\%, and 21.5\% at 14~T. Consequently, while the third channel is practically invisible at zero field, it becomes essential for describing the high-field curvature of $\rho_{xx}$ and $\rho_{xy}$. Due to its negligible weight at zero field, the parameters $n_{e2}$ and $\mu_{e2}$ are treated as effective quantities.

The formal compensation of all three channels is less robust than the compensation of the $e_1$--$h$ pair. The total compensation ratio, $C_{\mathrm{tot}}=n_h/(n_{e1}+n_{e2})$, is approximately 0.856, 0.984, 0.993, and 0.994 for the 110, 241, 301, and 783~nm samples, respectively. The largest deviation from perfect compensation occurs in the 110-nm flake, primarily due to the effective low-mobility $e_2$ channel, which contributes only approximately 0.09\% to the conductivity at $B=0$. Thus, the near-perfect compensation of the high-mobility $e_1$--$h$ pair serves as a more reliable physical descriptor.

The extracted carrier concentrations $n_i$ are also effective parameters; the multivalley Fermi surface of Sb, characterized by a tensor mobility, is simplified here into several scalar channels. The main results are the near-perfect compensation of the $e_1$--$h$ pair, the significant enhancement of $\mu_{\mathrm{pair}}$ with increasing thickness, and the emergence of a weak, low-mobility component in high magnetic fields.

\newpage
\subsection{Quantum Oscillations, Three-Dimensional Fermi Surface, and Size-Dependent Scattering}

While transport analysis provides effective carrier concentrations and mobilities, it does not provide direct information regarding the Fermi-surface geometry. This information is accessible through SdH oscillations: according to the Onsager relation, their frequency is proportional to the extremal orbit area \cite{ref18,ref19,ref20,ref27}. Consequently, quantum oscillations provide a direct test of whether the thirty-fold change in MR stems from a reconstruction of the Fermi surface or from variations in the scattering rate.

\begin{figure*}[t]
  \centering
  \includegraphics[width=0.96\textwidth]{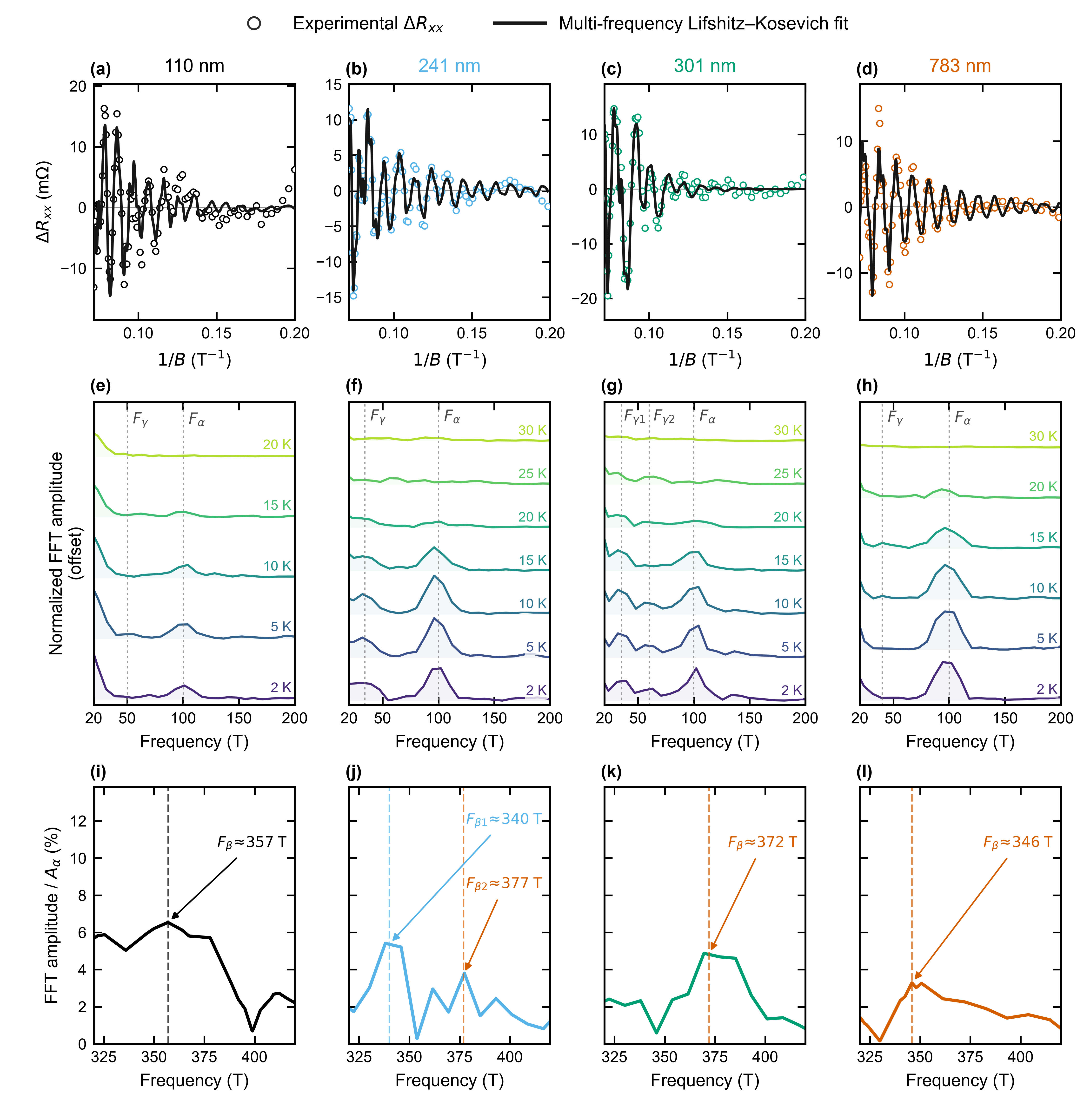}
  \caption{Thickness-dependent quantum oscillations in Sb flakes. (a)--(d) Oscillatory longitudinal resistance $\Delta R_{xx}$ at 2~K (symbols) and multifrequency Lifshitz--Kosevich fits (solid lines). (e)--(h) Temperature evolution of the vertically offset FFT spectra, highlighting the low-frequency $F_{\gamma}$ sector and the principal $F_{\alpha}$ orbit. (i)--(l) High-frequency FFT spectra revealing the bulk electron $F_{\beta}$ orbit.}
  \label{fig:sdh}
\end{figure*}

The oscillating component, $\Delta R_{xx}$, was extracted from the even part of the longitudinal resistance by subtracting a smooth magnetoresistive background. The analysis was restricted to the 5--14~T range, where the oscillations are reproducible and their amplitude significantly exceeds the low-field background. The data were interpolated onto a uniform $1/B$ grid prior to Fourier transformation. The frequency resolution is determined by the width of the reciprocal-field window:
\begin{equation}
\delta F\approx\left(\frac{1}{B_{\min}}-\frac{1}{B_{\max}}\right)^{-1}\approx7.8~\mathrm{T}.
\label{eq:fft_resolution}
\end{equation}
The stability of the frequencies was verified using an ensemble of plausible backgrounds and varying field windows. We considered only peaks that were reproducible in forward and reverse field sweeps, across the temperature and angular series, and in the profile analysis. The dispersion was quantified using the median and the median absolute deviation. Fundamental frequencies, harmonics, and valley-related contributions can overlap in Sb, making these reproducibility checks necessary. Figure~\ref{fig:sdh} shows the thickness-dependent quantum oscillations. The oscillatory waveform in reciprocal field was modeled using the multicomponent Lifshitz--Kosevich (LK) formalism:
\begin{equation}
\Delta R_{xx}=\sum_j A_{0,j}R_{T,j}(B)R_{D,j}(B)
\cos\left[2\pi\left(\frac{F_j}{B}+\phi_j\right)\right],
\label{eq:lk}
\end{equation}
\begin{equation}
R_{T,j}(B)=\frac{X_j(B)}{\sinh X_j(B)},\qquad
X_j(B)=\frac{2\pi^2k_BT m_j^{*}}{e\hbar B},
\label{eq:lk_thermal}
\end{equation}
\begin{equation}
R_{D,j}(B)=\exp\left(-\frac{2\pi^2k_BT_{D,j}m_j^{*}}{e\hbar B}\right).
\label{eq:lk_dingle}
\end{equation}
In the direct fitting of the oscillations, the local value of $B$ was used at each experimental data point. Only in the analysis of the temperature damping of the FFT amplitude within the 5--14~T window was the damping factor evaluated at $B_{\mathrm{eff}}=2/(B_{\min}^{-1}+B_{\max}^{-1})\approx7.37$~T. The multicomponent LK fits reproduce both the oscillation period and envelope with $R^2=0.81$--0.94; effective masses were extracted only for temperature-trackable maxima without switching between neighboring branches.

The most robust feature common to all four flakes is the $\alpha$ branch: $F_{\alpha}=98.7$, 96.1, 103.0, and 97.3~T for thicknesses of 110, 241, 301, and 783~nm, respectively. The average value is $98.8\pm3.0$~T; the relative standard deviation is approximately 3\%, and no monotonic thickness-dependent shift is observed. This is consistent with the known hole orbit of bulk Sb for a magnetic field close to the trigonal axis \cite{ref5,ref6,ref7,ref8}. From the Onsager relation,
\begin{equation}
A_{\mathrm{ext}}=\frac{2\pi eF}{\hbar},\qquad
k_{\mathrm{eq}}=\sqrt{\frac{A_{\mathrm{ext}}}{\pi}}.
\label{eq:onsager}
\end{equation}
The Onsager relation yields $A_{\alpha}=0.917$--0.983~nm$^{-2}$ and $k_{\mathrm{eq},\alpha}=0.540$--0.559~nm$^{-1}$. For the 241-nm flake, the damping of the component near 96~T gives $m^{*}\approx0.117m_e$, close to the known scale of the $\alpha$ orbit. In the following analysis, we use the conservative value $m_{\alpha}^{*}\approx0.10m_e$, corresponding to $v_F\approx(6.2$--$6.5)\times10^5$~m/s and $E_F\approx0.11$--0.12~eV.

Below $F_{\alpha}$, a $\gamma$ sector appears. In the 241-nm sample it lies near 41--47~T and has $m_{\gamma}^{*}=(0.045\pm0.006)m_e$; in the 301-nm sample, branches near 35 and 56--63~T are resolved, with $m_{\gamma}^{*}\approx0.064m_e$ for the higher-frequency branch. For the 110-nm flake, ensemble decomposition of the broad 50--70~T sector provides a tentative $m_{\gamma}^{*}\approx0.070m_e$; for the 783-nm flake, solutions that retain a weak internal maximum give $m_{\gamma}^{*}\approx0.044m_e$. The latter estimate exhibits some instability across different processing variants, and the frequency of the broad sector in the 110-nm flake depends on the local solution; therefore, both masses are marked as tentative in Table~\ref{tab:sdh} and are not used to calculate $v_F$ and $E_F$. It is more appropriate to refer to a sector of closely spaced low-frequency orbits rather than to a single line across the entire sample series.

For the most reliable $\gamma$ branches, similar energy scales are obtained: $v_F=9.0\times10^5$~m/s and $E_F\approx105$~meV for 241~nm, and $v_F=7.7\times10^5$~m/s and $E_F\approx107$~meV for 301~nm. This supports their association with the low-energy fermiology, but does not prove a common microscopic origin of the entire $\gamma$ sector.

\begin{table*}[t]
\caption{Main quantum-oscillation frequencies and effective transport scale $\ell_{\mathrm{eff}}$. Stable ranges are reported for the $\gamma$ sector; $\beta$ maxima are obtained after accounting for $\alpha$ harmonics. The dagger marks tentative effective LK estimates for broad or weak $\gamma$ features.}
\label{tab:sdh}
\begin{ruledtabular}
\begin{tabular}{c c c c c c}
$t$ (nm) & $F_{\gamma}$ (T) & $F_{\alpha}$ (T) & $F_{\beta}$ (T) & $m_{\gamma}^{*}/m_e$ & $\ell_{\mathrm{eff}}$ (nm) \\
\colrule
110 & 50--70, broad & 98.7 & 357.0 & $\approx0.070^{\dagger}$ & 447 \\
241 & 41--47 & 96.1 & 340; shoulder $\approx377$ & 0.045 & 765 \\
301 & $\approx35$; 56--63 & 103.0 & 372.0 & 0.064 & 972 \\
783 & 40--42, weak & 97.3 & 346.0 & $\approx0.044^{\dagger}$ & 2434 \\
\end{tabular}
\end{ruledtabular}
\end{table*}

The high-frequency $\beta$ sector was separated from $\alpha$ harmonics by explicitly including $F_{\alpha}$, $2F_{\alpha}$, $3F_{\alpha}$, and $4F_{\alpha}$ in the profile model. After this procedure, peaks at 361.0, 335.5, 377.0, and 352.5~T remain in all flakes. The mean value is $F_{\beta}=356.5\pm17.3$~T, and $F_{\beta}/F_{\alpha}=3.608\pm0.079$; the main peak lies below $4F_{\alpha}$ by $38.6\pm6.9$~T, exceeding the processing-induced variation.

The 335--377~T range corresponds to the electron orbit of the $L$-point pockets in bulk Sb \cite{ref5,ref7,ref8}. For this orbit, $A_{\beta}=3.20$--3.60~nm$^{-2}$ and $k_{\mathrm{eq},\beta}=1.01$--1.07~nm$^{-1}$. The shoulder near 382~T in the 241-nm flake coincides with $4F_{\alpha}$ and remains unresolved. The $\beta$ mass cannot be extracted from the available data; therefore, the assignment is based on frequency, reproducibility, and the harmonic test rather than on a complete reconstruction of the $L$ valleys.

Angle-dependent measurements were also performed. The angle $\theta$ is measured from the flake plane: $\theta=90^{\circ}$ corresponds to the magnetic field along the surface normal, $C_3\parallel c\parallel[001]$, whereas $\theta=0^{\circ}$ corresponds to an in-plane field. Two rotation series were performed in the mutually orthogonal experimental $C_1$--$C_3$ and $C_2$--$C_3$ planes. For a local triaxial ellipsoid, the extremal-section frequency is described by
\begin{equation}
F(\theta)=\left[
\frac{\cos^2\theta}{F_i^2}+\frac{\sin^2\theta}{F_{C3}^2}
\right]^{-1/2}.
\label{eq:angular}
\end{equation}
A joint fit of the $\alpha$ branch in the two rotation planes yields $F_{C1}=81.7$~T, $F_{C2}=72.5$~T, and $F_{C3}=96.7$~T, with a median deviation of 1.8~T. The corresponding effective semiaxes in the experimental frame, $k_{C1}=0.510$, $k_{C2}=0.576$, and $k_{C3}=0.431$~nm$^{-1}$, give a moderate anisotropy of $k_{\max}/k_{\min}=1.33$ and confirm the closed three-dimensional character of the $\alpha$ orbit.

For the $\gamma$ sector, the local approximation gives characteristic frequencies of 56.9, 34.3, and 42.5~T with an anisotropy of $k_{\max}/k_{\min}\approx1.66$. The estimated carrier density for a degeneracy of 3--6 is $(4.9$--$9.8)\times10^{18}$~cm$^{-3}$, i.e., of the same order of magnitude as the effective $e_2$ channel for the 241- and 301-nm flakes. This agreement in magnitude does not establish a unique correspondence between the orbit and a transport channel, nor does it determine the carrier sign. The frequencies remain finite for an in-plane magnetic field and do not follow a two-dimensional angular law; therefore, the $\gamma$ sector cannot be reduced to a simple two-dimensional surface orbit.

Quantum oscillations also allow the transport and quantum scattering scales to be compared. The quantum lifetime and mobility are obtained from the Dingle temperature as \cite{ref20,ref32}
\begin{equation}
\tau_q=\frac{\hbar}{2\pi k_BT_D},\qquad
\mu_q=\frac{e\tau_q}{m^{*}}.
\label{eq:dingle_parameters}
\end{equation}
For the $\alpha$ branch, median $T_D$ values of 13.5 and 17.6~K for the 241- and 301-nm flakes give $\tau_q=9.0\times10^{-14}$ and $6.9\times10^{-14}$~s, and $\mu_{q,\alpha}=1.36\times10^3$ and $1.21\times10^3$~cm$^2$\,V$^{-1}$\,s$^{-1}$, respectively. Since $\alpha$ is identified as a hole orbit, we compare $\mu_{q,\alpha}$ with the transport mobility $\mu_h$ obtained from the three-channel fit: the ratios $\mu_h/\mu_{q,\alpha}$ are approximately 15.4 and 24.8. These values demonstrate a substantial separation between the transport and quantum mobility scales, but they are not interpreted as exact band-resolved ratios of $\tau_{\mathrm{tr}}/\tau_q$.

To compare transport across thicknesses, we introduce an effective transport scale
\begin{equation}
\ell_{\mathrm{eff}}=\frac{\mu_{\mathrm{pair}}\hbar k_{\mathrm{eq},\alpha}}{e}.
\label{eq:leff}
\end{equation}
For thicknesses of 110, 241, 301, and 783~nm, this gives $\ell_{\mathrm{eff}}\approx447$, 765, 972, and 2434~nm, with $\ell_{\mathrm{eff}}/t\approx4.06$, 3.18, 3.23, and 3.11, respectively. Since $\mu_{\mathrm{pair}}$ is the geometric mean of the electron and hole transport channels, whereas $k_{\mathrm{eq},\alpha}$ corresponds to a specific hole $\alpha$ orbit, $\ell_{\mathrm{eff}}$ is used as a comparative transport scale rather than as a strictly band-resolved microscopic mean free path. The scaling $\ell_{\mathrm{eff}}\propto t^{0.87}$ is mathematically linked to $\mu_{\mathrm{pair}}\propto t^{0.87}$ for an almost constant $k_{\mathrm{eq},\alpha}$; it is consistent with dimension-limited transport relaxation, but does not constitute independent proof of boundary scattering.

\begin{figure*}[t]
  \centering
  \includegraphics[width=0.98\textwidth]{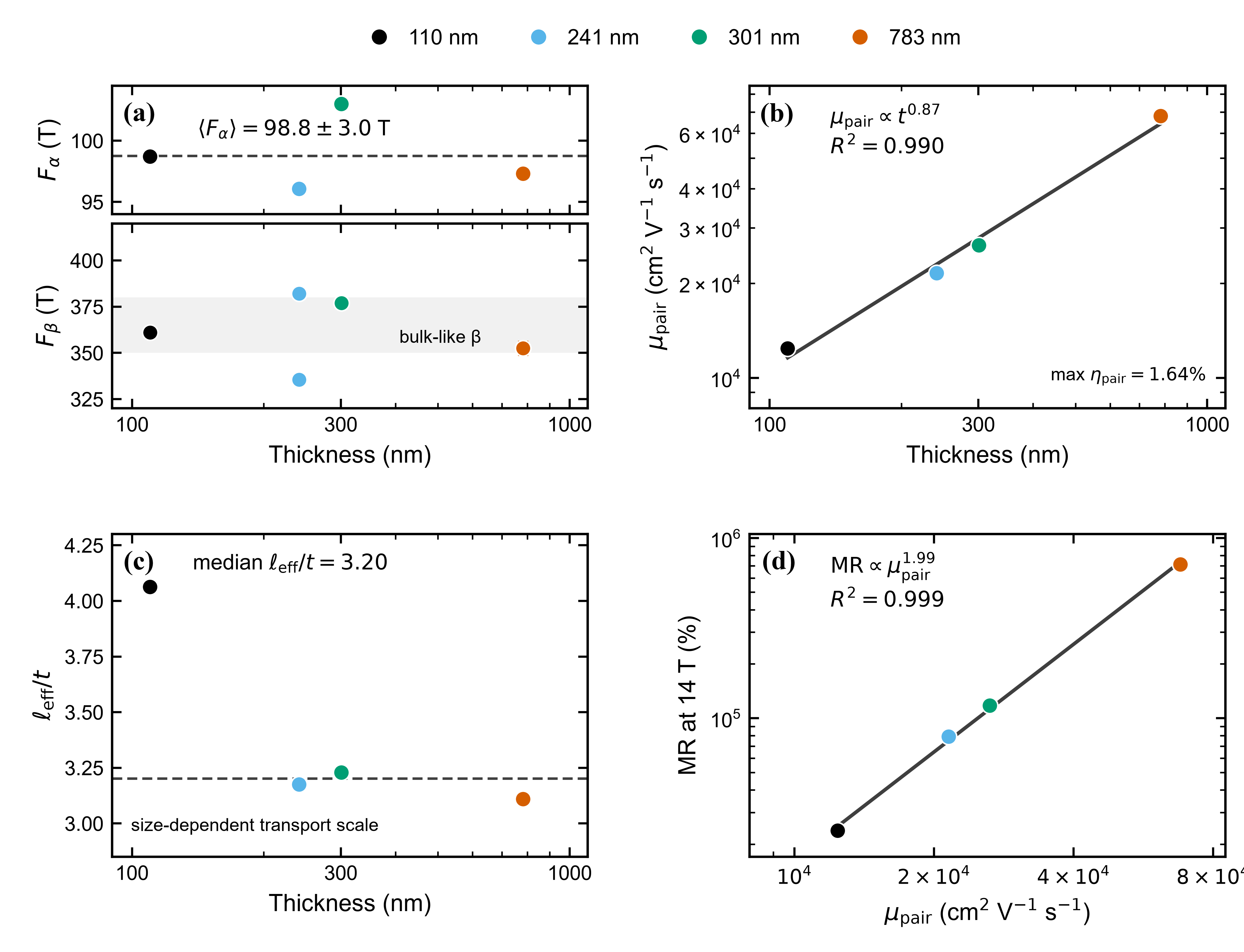}
  \caption{Thickness-dependent fermiology and transport relaxation in Sb flakes. (a) Frequencies of the principal $F_{\alpha}$ and bulk-electron $F_{\beta}$ orbits; the shaded region marks the bulk-like $\beta$ sector. (b) Mobility of the nearly compensated electron--hole pair as a function of thickness. (c) Effective transport length normalized by the flake thickness. (d) Magnetoresistance at 14~T as a function of $\mu_{\mathrm{pair}}$. Solid lines show power-law fits, whereas dashed lines indicate the corresponding mean or median values.}
  \label{fig:integrated}
\end{figure*}

Figure~\ref{fig:integrated} summarizes the thickness dependence of the SdH frequencies and transport parameters. The relative standard deviation of $F_{\alpha}$ is approximately 3\%, and $F_{\beta}$ exhibits no monotonic drift, whereas $\mu_{\mathrm{pair}}$ increases by roughly a factor of 5.5 and $\mathrm{MR}(14~\mathrm{T})$ by roughly a factor of 30. Quantitatively, $\mu_{\mathrm{pair}}\propto t^{0.87}$, while the experimental $\mathrm{MR}(14~\mathrm{T})$ scales almost quadratically as $\mu_{\mathrm{pair}}^{1.99}$ ($R_{\log}^2\approx0.999$). This is consistent with a nearly compensated orbital mechanism controlled by thickness-dependent transport relaxation; the third channel remains necessary for describing the full field and Hall response.

The light $\gamma$ sector becomes more pronounced in thinner samples and cannot be reduced to a simple two-dimensional surface orbit. Thickness modifies the transport lifetime and spectral weights, but does not induce a monotonic reconstruction of the main extremal cross sections.

\section{Discussion}

The different thickness dependences of the SdH frequencies and transport mobility point to a change in relaxation rather than a reconstruction of the primary extremal Fermi-surface areas. The effective length scale $\ell_{\mathrm{eff}}$ remains on the order of 3.1--4.1 times the flake thickness. Since $\ell_{\mathrm{eff}}$ is derived from $\mu_{\mathrm{pair}}$ and a nearly constant $k_{\alpha}$, this ratio expresses the same thickness dependence rather than providing independent verification. A natural explanation is the reduction of boundary and surface scattering with increasing $t$; however, the current measurements do not uniquely determine the microscopic surface specularity or the individual Fuchs--Sondheimer mean free path.

Classical transport and SdH oscillations weight the carrier contributions differently. In the zero-field limit, the contribution of a channel to $\sigma_{xx}$ scales as $n\mu$, whereas the low-field Hall response is weighted by $s_i n_i\mu_i^2$; in high fields, both weights are suppressed by the factor $[1+(\mu B)^2]^{-1}$. In contrast, SdH oscillations selectively probe orbits with sufficiently long quantum lifetimes. Consequently, the correspondence between the $\gamma$ sector and the weak $e_2$ channel should be interpreted only as a plausible matching of carrier concentrations; the data do not strictly establish the carrier sign, valley degeneracy, or the unambiguous assignment of the orbit to a specific pocket.

The discrepancy between the hole mobility $\mu_h$ from the three-band fit and the quantum mobility $\mu_{q,\alpha}$ reflects the different sensitivity of the transport and quantum lifetimes to the scattering angle \cite{ref25,ref32}. Since Dingle parameters are sensitive to the frequency window and $\mu_h$ is an effective scalar parameter within a multivalley model, the ratios $\mu_h/\mu_{q,\alpha}\approx15$--25 should be viewed as characteristic scales rather than precise measures of the microscopic $\tau_{\mathrm{tr}}/\tau_q$ ratio.

\section{Conclusion}

In CVD-grown Sb flakes (110--783~nm), the primary quantum-oscillation orbits ($\alpha$ and $\beta$) retain their bulk-like character, while the transport mobility of the nearly compensated high-mobility pair scales as $t^{0.87}$. A weak third channel, potentially associated with the $\gamma$ sector, remains necessary to describe the full conductivity tensor. The observed near-quadratic scaling of the magnetoresistance, $\mathrm{MR}(14~\mathrm{T})\propto\mu_{\mathrm{pair}}^{1.99}$, provides evidence that the strong thickness dependence of MR is driven by size-dependent transport relaxation rather than a systematic reconstruction of the primary extremal Fermi-surface sections.

\begin{acknowledgments}
This work is supported by the National Key R\&D Program of China (Grant No.~2022YFA1203902), the National Natural Science Foundation of China (NSFC) (Grant Nos.~12374108 and 12241401), the Guangdong Provincial Quantum Science Strategic Initiative (Grant Nos.~GDZX2401002 and GDZX2501003), the GJYC program of Guangzhou (Grant No.~2024D01J0087), the Fundamental Research Funds for the Central Universities (No.~2232026D49), and the Shanghai Partner Research Program of the Science and Technology Commission of Shanghai Municipality (Grant No.~26HB2700800).
\end{acknowledgments}

\section*{Author Contributions}
M.G. and R.W. conceived the study, designed the experiments, analyzed the data, and interpreted the results. M.G. performed flake growth and device fabrication. M.G. and J.W. performed magnetotransport measurements. M.G. prepared the initial figures and first draft of the manuscript. R.W. supervised the project, provided resources, and contributed to methodology development and result validation. L.Z. performed FIB specimen preparation and HAADF-STEM characterization. All authors discussed the results, reviewed the manuscript, and approved the final version.

\section*{Data Availability}
The data supporting the findings of this study are available from the corresponding author upon reasonable request.

\bibliography{references}

\end{document}